\pdfoutput=1
\documentclass[letterpaper, 10 pt, conference]{ieeeconf}
\usepackage{graphicx}
\usepackage{xcolor}
\usepackage{amsmath}
\usepackage{ragged2e}
\usepackage{breqn}
\usepackage{caption}
\DeclareCaptionLabelSeparator{none}{ }
\usepackage{authblk}
\usepackage{hyperref}
\usepackage{dblfloatfix}
\usepackage{lipsum}
\usepackage{fancyhdr}
\usepackage{lastpage}
\usepackage{multicol}
\usepackage{longtable}
\usepackage{float}

\title{Tethers_Paper1}

\title{\center\\\textbf{
\large Optimum Location to Intercept Interstellar Objects with Build-and-wait Missions}
}
\begin{document}
\pagestyle{fancy}
\renewcommand\Authfont{\small}
\renewcommand\Affilfont{\itshape\footnotesize}
\vspace{-10mm}


\twocolumn[{
\begin{center} 
\vspace{5mm}
\textbf{\large Optimum Location to Intercept Interstellar Objects with Build-and-wait Missions}\\
\vspace{5mm}
\textbf{Laia López Llobet$^{a}$*},  \textbf{Andreas M. Hein$^{b}$} 
\\
\end{center}

$^{a}$ \emph{International Space University, 1 rue Jean-Dominique Cassini, 67400 Illkirch-Graffenstaden, France,} \\ 
\underline{laia.lopez.llobet@community.isunet.edu} \\
$^{b}$ \emph{Initiative for Interstellar Studies, 27/29 South Lambeth Road, SW8 1SZ London, United Kingdom,} \\ \hspace{1.5mm}
\underline{andreas.hein@i4is.org} \\ 
 * Corresponding Author

\begin{center}
\textbf{Abstract}
\end{center}

Until now, only two Interstellar Objects (ISOs) have been discovered: 1I/'Oumuamua and 2I/Borisov. Despite a limited amount of observations, they present high scientific interest for the research community. In order to further analyze these objects, in-situ data is required. However, beginning the mission design and spacecraft manufacture after ISO detection would significantly lower mission success probability. Consequently, “build-and-wait" is an appealing mission concept.
To prepare the upcoming missions to ISOs it is required to have a deeper knowledge on the probability analysis to locate these objects within the Solar System. Therefore, the optimum location to intercept these objects has been studied by performing a computational analysis using the Runge-Kutta RKF4(5) numerical method. This method has been used to determine the likelihood of ISOs trajectories intercepting the ecliptic. The results show a peak at a distance of $\sim$1 AU from the Sun. Therefore, the vicinity of the Earth’s orbit is the best position to send or place a spacecraft for ISO interception. In addition, these results add validity to the theory of panspermia via ISO impact. The author has estimated an annual probability of ISO-Earth impact of 10$^{-7}$.

\hspace{0.5mm} \textbf{Keywords:} Build-and-wait, Interstellar Objects, Probability Analysis, Space Missions\\
\vspace{5mm}
}]

\thispagestyle{fancy}



\section{\textbf{1. Introduction}}
\justifying
Since the detection of the first interstellar objects (ISOs\footnote{\textbf{ISO}: Interstellar Object}) passing through the Solar System, the scientific community has been working on different approaches to study them. The first ISOs ever detected were 1I/'Oumuamua in October 2017 \cite{Meech}, followed by 2I/Borisov in August 2019 \cite{Guzik}. They were classified as interstellar asteroid and interstellar comet, respectively. Such ISOs present high scientific interest because they could provide more insight into the composition of other stellar systems, planetary formation and the hypothesis of panspermia, among others.

Although astronomical observations of ISOs can be performed by telescope, in order to further analyze these objects, in-situ data gathering by spacecraft is required. Such objects are usually detected late in their passage through the Solar System and possess high heliocentric velocities. Their detection frequency is low and the observation window is typically short. These obstacles make both reaching these scientifically promising objects and also designing an appropriate spacecraft challenging. For this reason, mission design assessment is fundamental for the development of the first missions to ISOs.

Different feasibility studies have been carried out and have stated that missions to ISOs are possible with current technology. At least two possible missions to 1I/’Oumuamua with launch dates between 2030 and 2033 and arrival dates in 2048 and 2052 have been identified \cite{Hibberd1}. Furthermore, Hibberd et al. \cite{Hibberd2} determined that the interstellar comet 2I/Borisov could be reached in 2045 if a mission was launched in 2030. Hein et al. \cite{Hein} present three different approaches: flybys with an impactor, flybys with returning samples and rendezvous missions with an orbiter or lander. An example of a flyby with kinetic impactors is the Bridge concept developed by Moore et al. \cite{Moore}, a mission approach for ISO in-situ exploration where a spacecraft releases an impactor and performs a deflection maneuver. 

However, sending a spacecraft to known ISOs that are currently exiting the Solar System is not the only mission approach that has been proposed. Another option for intercepting ISOs is to build a spacecraft (or multiple spacecraft) for future detections. These types of missions are referred to as "build-and-wait" missions. The spacecraft can be kept on Earth or be sent to a specific position in space to wait for its target. Such types of missions will be supported by the new Large Synoptic Survey Telescope that will start surveying in 2022 and could be able to detect one ISO per year, according to Trilling et al. \cite{Trilling}.

The start of the first missions to ISOs is imminent and very promising. However, it is still a topic not very much studied that needs more insight and further assessment. For this reason, this paper aims to provide an assessment of the best location to intercept ISOs with build-and-wait space missions.

The structure of the paper is the following: Section 2 presents the study to determine the optimum location of ISO interception and Section 3 includes the project results and discussion. Finally, conclusions of the project are presented in Section 4, followed by an appendix with a trade-off between build-and-wait missions on Earth and in space.

\section {\textbf{2. Optimum Location for the Interception of Interstellar Objects}}
This chapter describes the study to analyze the trajectories of ISOs entering the Solar System and determine the optimum location for ISO interception in the vicinity of the Earth. 

\subsection{2.1 Methodology}
\begin{figure}[H]
    \centering
\includegraphics[width=0.47\textwidth]{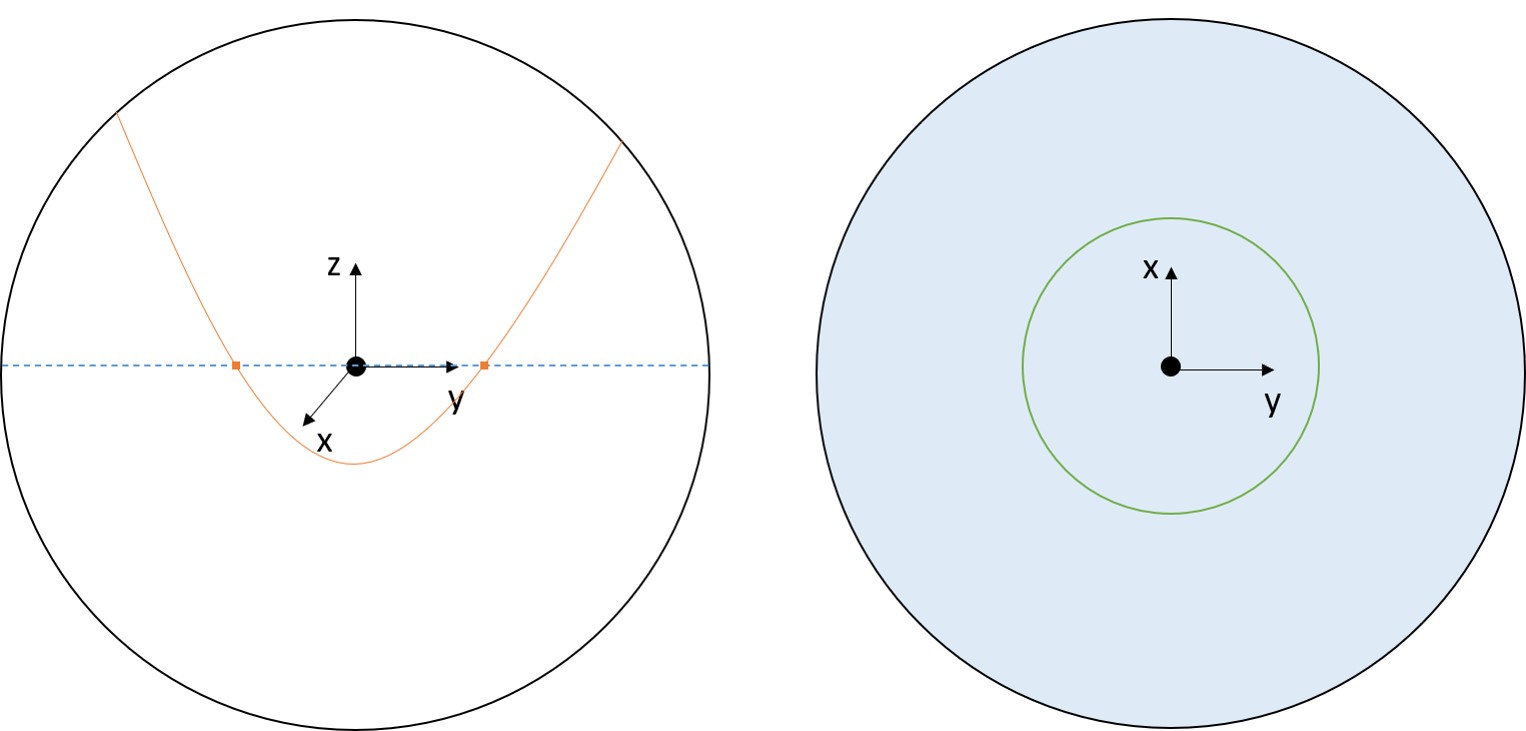}
  \caption{Representation of the method followed to calculate the optimum location for ISOs’ interception. The figure on the left represents the volume of the Solar System studied. The black dot represents the Sun, the blue dashed line represents the ecliptic, the orange line represents the ISO trajectory and the orange squares represent the ISO-ecliptic intersection. The figure on the right shows another representation of the volume studied. Color blue represents the ecliptic plane and the circle in green represents a circular orbit centered in the Sun. The coordinate system’s axes have been represented for a better understanding of the images.}
  \label{fig:method}
\end{figure}

The main objective is to simulate the potential trajectories that ISOs might follow in the Solar System and calculate the intersection point between the ecliptic plane and the ISO trajectory. This is performed to characterize the ISOs’ trajectories in the inner Solar System and determine the likelihood of ISOs-ecliptic intersection within the Solar System. Therefore, this analysis shows whether the Earth is a good location to launch direct build-and-wait missions to ISOs or not. In addition, the results of this section give more insight on build-and-wait missions in space too, since the best position to place a spacecraft is identified. Fig. \ref{fig:method} shows graphically the method used and illustrates the intersection of the ISO’s trajectory and the ecliptic.

When talking about space missions, fuel availability is one of the major constraints. For this reason, spacecraft take advantage of the gravity pull from other planets to perform maneuvers and gravity assists to their mission trajectory. Due to these energetic constraints, the optimal interception location should be near the plane of the orbits of the Solar System’s planets. In addition, it should be as close to Earth as possible, in order to decrease the complexity of the mission and the amount of fuel required.

\begin{figure}[H]
    \centering
\includegraphics[width=0.3\textwidth]{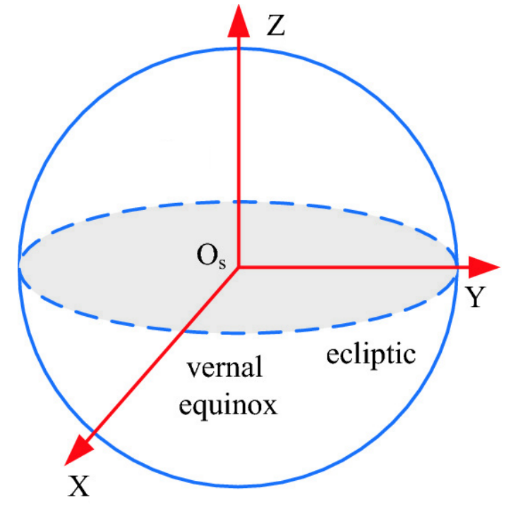}
  \caption{Heliocentric-ecliptic coordinate system. The origin of coordinates (Os), fundamental plane (ecliptic) and reference direction (vernal equinox) are represented in cartesian coordinates \cite{Wang}.}
  \label{fig:coordinates}
\end{figure}

To treat this problem, the heliocentric-ecliptic coordinate system is used (see Fig. \ref{fig:coordinates}), with its origin at the center of the Sun, the ecliptic plane as the fundamental plane and the vernal equinox as the reference direction. The ecliptic plane has been chosen as the reference plane, from which most planets’ orbits stay within 3 degrees, except for Mercury which orbital plane is inclined 7 degrees with respect to the ecliptic.

In order to simplify calculations, the problem is treated as a two-body problem in three dimensions, only taking into account the Sun and the ISO. The ISOs’ trajectories have been created by a MATLAB program written by the author, which generates ISOs’ trajectories from the ISO initial position and velocity.

\subsection{2.2 Setting the Variables}
The first step of the study is the assignment of mass, initial position and velocity vectors for the Sun and the ISOs, according to the coordinate system chosen. 

\subsection{2.2.1 Mass}

The value for the Sun’s mass from \cite{Williams} is used. For simplicity, the ISO has been assumed to be spherical and its mass has been calculated accordingly. The density value used corresponds to the average asteroid density estimation in the Solar System. Even though the asteroid density varies with the asteroid spectral type, in general, a value of $\rho_2$= 2000 kg/m$^3$ is assumed. Following the spherical assumption, the volume has been calculated using $V_2$= $\frac{4}{3}\pi r_2^3$, where the author has assumed $r_2 \simeq 0.5$ km, according to the estimations for the radius of 1I/’Oumuamua and 2I/Borisov. Following this argument, the mass estimated for the ISOs is
$M_2\simeq 10^{12}$ kg.

\subsection{2.2.2 Initial Velocity Vectors}
\begin{figure}[H]
    \centering
\includegraphics[width=0.48\textwidth]{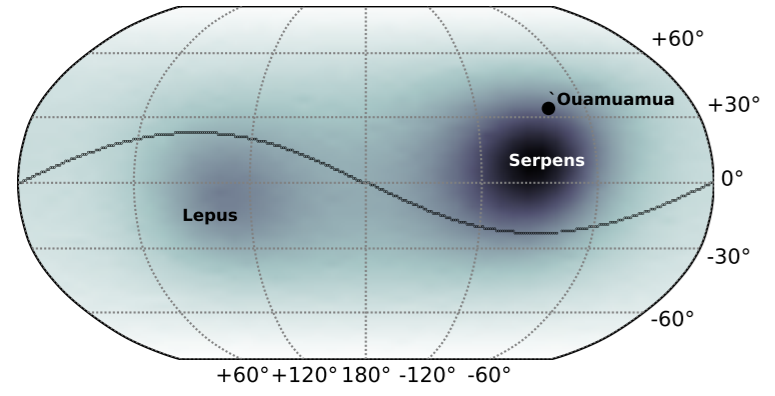}
  \caption{Probability map from a heliocentric perspective representing the likelihood of ISOs approaching with a velocity vector parallel to the vector pointing from the Sun to a particular position in the sky. A higher probability is indicated with darker colors. The ecliptic is represented by a black line. The positions of 1I/’Oumuamua entering the Solar System and the constellations Serpens and Lepus have also been represented. These constellations are in the solar apex and anti-apex, respectively \cite{Seligman}.}
  \label{fig:seligman}
\end{figure}

Since the Sun is the origin of the coordinate system, it remains at rest. When generating the initial velocity vectors for the ISOs, two different parameters have been taken into account: the velocity vector direction and magnitude. 
It is well known that the Sun travels at a certain velocity with respect to its local neighborhood in the Milky Way, known as the local standard of rest. For this reason, due to the movement of the Sun in the galaxy, there are certain directions from which ISOs might enter the Solar System with a higher likelihood. The author wanted to take into account this factor and, in order to do so, has used the probability analysis of  Seligman and Laughlin \cite{Seligman} depicted in Fig. \ref{fig:seligman}. This figure represents the probability of an ISO approaching on a trajectory parallel to the vector pointing from the Sun to that location of the sky. As can be observed in Fig. \ref{fig:seligman}, the solar apex direction presents a higher likelihood. 
The author has programmed the ISOs’ trajectories to have an initial velocity direction within the range of higher likelihood shown in Fig. \ref{fig:seligman}. From the graph, the range of values of the coordinates with higher likelihood is approximately $\alpha=(-120^\circ,-60^\circ)$, and $\delta=(-10^\circ,30^\circ)$.

As mentioned by \cite{Seligman}, these coordinates do not indicate the region of the sky from which it is more likely that ISOs will enter the Solar System, but rather indicate the velocity directions that the ISOs are more likely to present.
The program that the author has written generates random equatorial coordinates from the heliocentric perspective within the values with higher likelihood mentioned above. After the random generation of these coordinates for an ISO, the coordinates are transformed to cartesian coordinates $(X_0,Y_0,Z_0)$ by using the following formulas:
\begin{equation}
    X_0=r\cdot \cos\alpha \cdot \cos \delta
\end{equation}

\begin{equation}
    Y_0=r \cdot \sin \alpha \cdot \cos\delta
\end{equation}

\begin{equation}
    Z_0=r \cdot \sin \delta
\end{equation}

Where $\vec{r}$ is calculated by:

\begin{equation}
    \vec{r}=\vec{R}_{1,0}-(X_0,Y_0,Z_0 )=-(X_0,Y_0,Z_0 )
\end{equation}


Therefore, the unitary direction can be calculated by $\vec{R}_{unit} \frac{\vec{r}}{|\vec{r}|}$.

\begin{figure}[H]
    \centering
\includegraphics[width=0.48\textwidth]{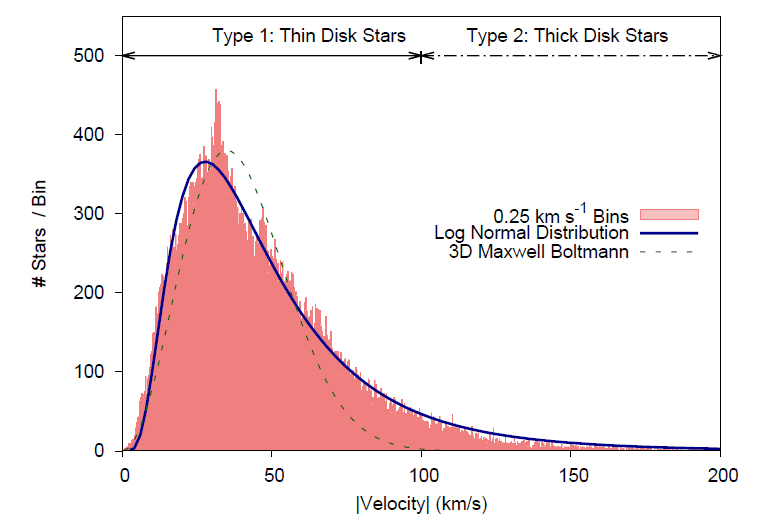}
  \caption{Histogram representing the stellar velocity magnitudes relative to the Solar System from the stellar population of thin and thick disk stars. This data has been obtained from the Gaia Early Data Release 3. The Log-Normal and Maxwell-Boltzmann distributions are represented in the figure for a better comparison with the histogram \cite{Eubanks2021}.}
  \label{fig:eubanks}
\end{figure}

Once the unitary direction of the vector has been determined, the next topic to be addressed is the magnitude of the velocity vector. The magnitude of v$_\infty$ that ISOs present is known. However, due to the gravity pull of the Sun, its velocity increases while the ISO is approaching the Sun. The values that the velocity of the ISO can take depends on different factors, such as the distance from the Sun and the eccentricity of the orbit. 
Since studies on the values of v$_\infty$ that ISOs are expected to present exist, the author has worked with the fact that at $\vec{R}_{2,0}$ the ISO presents a velocity equivalent to v$_\infty$. 1I/’Oumuamua and 2I/Borisov present v$_\infty \simeq$ 25 km/s and v$_\infty \simeq$ 32 km/s, respectively. However, a study on the expected  v$_\infty$ has been performed by Eubanks et al. \cite{Eubanks2021} (see Fig. \ref{fig:eubanks}). According to \cite{Eubanks2021}, “half of the predicted ISO arrivals will have velocities $\geq$ 38 km/s, and 50\% of the arrivals are predicted to fall between 22.5 and 62.5 km/s”.  

The values for the velocity magnitude have been generated following the Log-Normal distribution fitted into Fig. \ref{fig:eubanks}, which has the following parameters: $\mu_{LN}$=3.715 $\pm$ 0.003 and $\sigma_{LN}$=0.624$\pm$0.002 \cite{Eubanks2021}. 
Finally, the velocity has been calculated by the following formula:

\begin{equation}
    \vec{v}_{2,0}=|v_{LN}|\cdot \vec{R}_{unit}
\end{equation}

\subsection{2.2.3 Initial Position Vectors}
After using the probability analysis of Seligman and Laughlin \cite{Seligman} and Eubanks et al. \cite{Eubanks2021} to generate the initial velocity vector of ISOs, the ISO’s initial position needs to be considered. The main issue that needs to be addressed in this subsection is at what distance from the Sun the initial position of the ISO is set. 
As it has been mentioned in Subsection 2.2.2 , the initial ISO velocity has been considered to be v$_\infty$. Therefore, the initial ISO position needs to be far enough from the Sun in order for the initial velocity to be approximately v$_\infty$. To determine at which distance from the Sun the velocity can be approximated to v$_\infty$, the author has performed a parallel analysis in which the change in velocity measured as a percentage is plotted against the distance from the Sun. Fig. \ref{fig:simulation} shows an example of the results for 50 generated trajectories:

\begin{figure}[H]
    \centering
\includegraphics[width=0.48\textwidth]{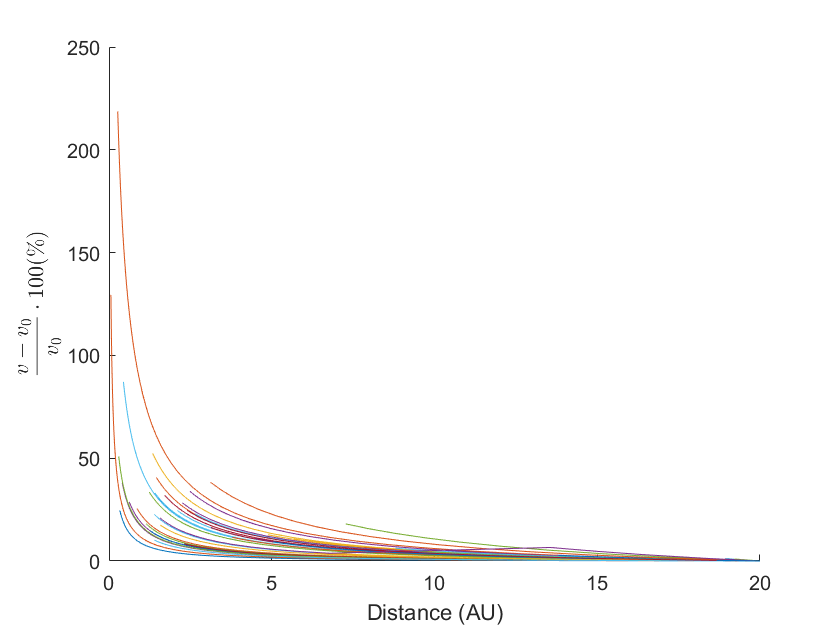}
  \caption{Representation of the change in velocity (in percentage) that the ISOs experience while approaching the Sun, as a function of the distance.}
  \label{fig:simulation}
\end{figure}

As it can be observed from Fig. \ref{fig:simulation}, the difference between the velocity of the ISOs and its initial velocity v$_0$, which we consider to be v$_\infty$, can be almost negligible at a distance of 10 AU. For this reason, the initial position of the ISOs has been generated randomly on the surface of a sphere with a radius of 10 AU, centered in the Sun.

\subsection{2.3 Generation of ISOs trajectories}

After having generated the initial velocity and position of ISOs, the trajectory is generated by taking into account the acceleration that the two bodies experience due to each other's gravity force. This acceleration is calculated using the following equations \cite{Curtis}:

\begin{equation}
\left( \ddot{X}_1, \ddot{Y}_1, \ddot{Z}_1 \right) = \frac{G_c M_2}{d^3} \left( X_2-X_1, Y_2-Y_1, Z_2-Z_1 \right) 
\end{equation}

\begin{equation}
\left( \ddot{X}_2, \ddot{Y}_2, \ddot{Z}_2 \right) = \frac{G_c M_1}{d^3} \left( X_1-X_2, Y_1-Y_2, Z_1-Z_2 \right) 
\end{equation}

Where $d=\sqrt{(X_2-X_1 )^2+(Y_2-Y_1 )^2+(Z_2-Z_1 )^2}$ and $G_c=6.67259\cdot 10^{-20}$ km$^3$/kg/s$^2$.
The Runge-Kutta RKF4(5) method is used in order to integrate the differential equations. This method, developed by Fehlberg, consists of a mix of RK4 and RK5 methods. It is known that using a constant step to integrate the equations can be inefficient and other methods exist to automatically adjust the step size. The RKF4(5) method consists of using two adjacent-order Runge-Kutta in one. This way, the difference between the higher and lower order solution is calculated to estimate the error and adjust the step size. 
The process has 6 stages and the derivatives are the following \cite{Curtis}:

\begin{equation}
 \mathbf{\tilde{f}_1}  = \mathbf{\tilde{f}}(t_i,\mathbf{y_i})
\end{equation}

\begin{equation}
\mathbf{\tilde{f}_2}  = \mathbf{\tilde{f}}(t_i + a_2 h,\mathbf{y_i}+hb_{21}\mathbf{\tilde{f}_1}) \\
\end{equation}

\begin{equation}
\mathbf{\tilde{f}_3}  = \mathbf{\tilde{f}}(t_i + a_3 h,\mathbf{y_i}+h[b_{31}\mathbf{\tilde{f}_1}+b_{32}\mathbf{\tilde{f}_2}]) \\
\end{equation}

\begin{equation}
\mathbf{\tilde{f}_4} = \mathbf{\tilde{f}}(t_i + a_4 h,\mathbf{y_i}+h[b_{41}\mathbf{\tilde{f}_1}+b_{42}\mathbf{\tilde{f}_2}+b_{43}\mathbf{\tilde{f}_3}]) \\
\end{equation}

\begin{equation}
\begin{split}
  \mathbf{\tilde{f}_5} = & \mathbf{\tilde{f}}(t_i + a_5 h,\mathbf{y_i}+h[b_{51}\mathbf{\tilde{f}_1}+b_{52}\mathbf{\tilde{f}_2} +b_{53}\mathbf{\tilde{f}_3}\\ &+b_{54}\mathbf{\tilde{f}_4}]) 
\end{split}
\end{equation}

\begin{equation}
\begin{split}
\mathbf{\tilde{f}_6} = & \mathbf{\tilde{f}}(t_i + a_6 h,\mathbf{y_i}+h[b_{61}\mathbf{\tilde{f}_1}+b_{62}\mathbf{\tilde{f}_2}+b_{63}\mathbf{\tilde{f}_3} \\ & +b_{64}\mathbf{\tilde{f}_4}+b_{65}\mathbf{\tilde{f}_5}]) 
\end{split}
\end{equation}

This method returns the position and velocity vectors integrated over time, from which can be obtained the ISO trajectory. Once the trajectory is obtained, the point intersecting the ecliptic can be easily determined. 
From the trajectories, the ISO eccentricity can be calculated by the following formula: 

\begin{equation}
    e=\bigg|\frac{1}{\mu} [(v^2-\frac{\mu}{|\vec{r}|})\vec{r}-rv_r \vec{v}]\bigg|
\end{equation}

Where $\mu$ follows the expression $\mu=G_c (M_1+M_2)$. Furthermore, $v_r$ is the radial velocity that has been calculated by $v_r=\frac{( \vec{r}\cdot \vec{v})}{|\vec{r}|}$.

ISOs are expected to present hyperbolic eccentricities (e\textgreater1). The author has also researched different studies regarding the distribution of expected eccentricities that ISOs might present. According to the analysis of the ISOs eccentricities distribution in an observable sphere of 12 AU, performed by Marčeta and Novaković \cite{Marceta}, the author has assumed an upper limit of e=9 to decrease the running time of the MATLAB program.

\section{\textbf{3. Results and Discussion}}

\begin{figure}[H]
    \centering
\includegraphics[width=0.38\textwidth]{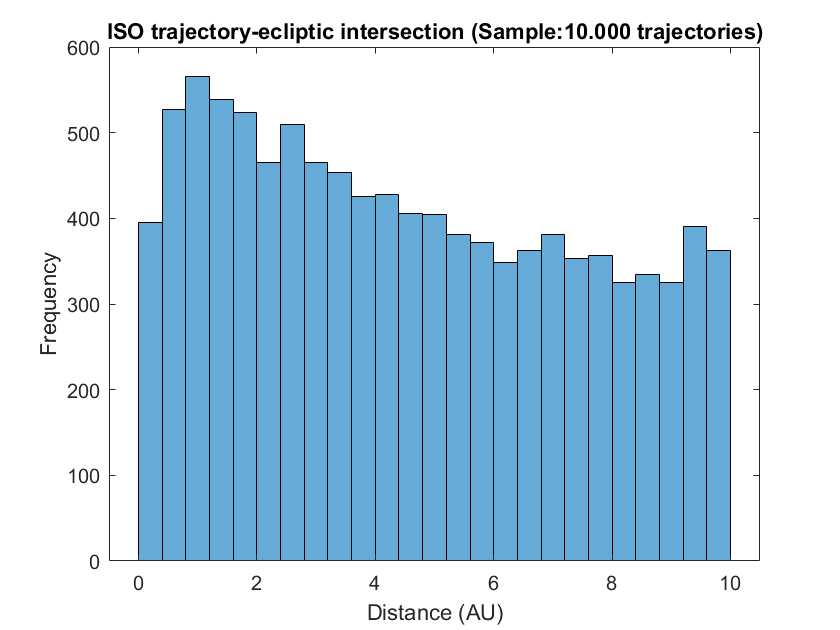}
  \caption{Histogram representing the frequency of ISOs intercepting the ecliptic at a given distance from the Sun with data obtained from 10.000 trajectories. 25 bins, each with a length of 0.4 AU, are represented.}
  \label{fig:histogram}
\end{figure}

The results of this study are shown in Fig. \ref{fig:histogram} where the frequency of ISO trajectory-ecliptic intersection can be observed. This graph has been plotted from a sample of 10.000 ISOs' trajectories.

The histogram peak, which represents the distance at which ISOs’ trajectories intercept the ecliptic with the highest likelihood, is in the range of $\sim$ 0.3-2 AU from the Sun, with a peak at approximately the position of the Earth (1 AU). Hence, these results show how the vicinity of the Earth orbit is the best position to place or send a spacecraft for ISO interception. In addition, it shows how the Earth is in a good position to detect ISOs and opens the door to the thought of prebiotic molecules being brought to Earth by the impact of an ISO.
Due to these results, a high-level probability analysis has been performed regarding the impact probability between an ISO and the Earth. 

It is assumed that the cross-section of the Earth is:
\begin{equation}
\sigma_{Earth}=\pi \cdot R_{Earth}^2\sim10^{14}  m^2
\end{equation}

Where R$_{Earth}\sim$6400 km and the cross-section within 1 AU of the Sun is estimated to be $\sigma_{1AU} \sim 10^{22}  m^2$.
Therefore, the Earth occupies 10$^{-8}$ of that surface area. 
According to \cite{Eubanks2021}, approximately 10 ISOs per year hit the Solar System plane within 1 AU of the Sun, assuming the number density of 1I/’Oumuamua objects. Therefore, with this assumption, the annual probability of an ISO hitting the Earth is 10$^{-7}$. This means that, on average, one ISO would hit the Earth every ten million years. It should be noted, however, that the fraction of ISOs with potential life-forms or prebiotic molecules is unknown and has not been taken into account. If this fraction is substantial, it opens the door to the hypothesis that the Earth could be the rare case where life was transmitted via an ISO. It should also be noted that this calculation has been made for objects with the size of 1I/’Oumuamua. However, there could be smaller ISOs in the Solar System that humans are unable to detect with the current detection capabilities.

\subsection{3.1 Surface Area Influence}

From Fig. \ref{fig:histogram}, it would be expected a decrease in frequency when farther away from the Sun. However, it should be noted that the surface area of the circles associated with the bin width increases as a square function with the distance. Therefore, the width of the bins corresponds to a radial section with a surface area of $\pi(r_{outer}^2-r_{inner}^2)$. 

\begin{figure}[H]
    \centering
\includegraphics[width=0.48\textwidth]{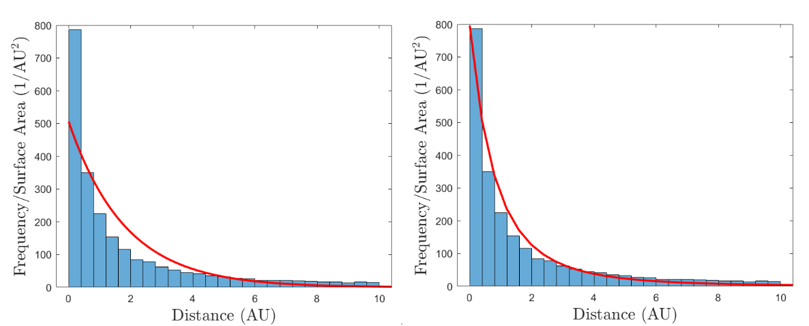}
  \caption{Exponential (left) and generalized Pareto (right) distribution function fit for the histogram representing the frequency divided by the surface area of ISOs intercepting the ecliptic at a given distance from the Sun. The sample analyzed consists of 10.000 trajectories.}
  \label{fig:histogram2}
\end{figure}

In addition, it should be taken into account that the placement of a spacecraft for ISO interception at a larger distance such as 9-10 AU is not optimum. This is due to the fact that even though the ISO would intercept the ecliptic in the same orbit as the spacecraft, the orbit would be so large that the time required for the spacecraft to travel from its position to the ISO would endanger mission success. In order to take this into account for the optimization of the ISO interception, the results obtained in Fig. \ref{fig:histogram} have been scaled considering the surface area difference. These results are represented in Fig. \ref{fig:histogram2}.

Different distribution functions \cite{Matlab} have been fitted in the histogram with 10.000 trajectories to determine the distribution function that the data presents. The exponential and generalized Pareto distribution functions are the most accurate. 

This section concluded that the likelihood of ISO’s intersection with the ecliptic is higher in the range of $\sim$0.3-2 AU from the Sun, with a peak at $\sim$1 AU. It can be concluded that near the Earth orbit is the best position to place a spacecraft for ISO interception. In addition, these results open the door to the thought of life being brought to Earth by the impact of an ISO. The annual probability of an ISO hitting the Earth has been calculated to be 10$^{-7}$. The calculation was made without taking into account the fraction of ISOs that could carry prebiotic molecules.

\section{\textbf{4. Conclusions}}

This paper presented the current knowledge of ISOs and determined, by the RKF4(5) numerical method using MATLAB, the distance from the Sun at which ISOs’ trajectories are more likely to intersect the ecliptic.

The best orbital approach to ISOs is a fast flyby with kinetic impactors, due to the high relative velocity expected between the spacecraft and the object. The kinetic impactors would collide with the ISO, allowing to obtain data of both surface and subsurface material, and the volatile characterization of the object in case of outgassing. 

The determination of the optimum location to intercept an ISO in the Solar System has been performed using the RKF4(5) numerical method. The intersection between the ISO trajectory and the ecliptic plane has been concluded to be the optimum location due to energetic constraints. By assuming a two-body problem, different estimations have been performed regarding the density and size of ISOs. The initial ISO velocity vector of the ISO trajectory simulated has been determined taking into account the probability analysis of Seligman and Laughlin \cite{Seligman} and the ISO analysis performed by Eubanks et al. \cite{Eubanks2021}. The initial ISO position has been determined randomly on the surface of a sphere with a radius of 10 AU.

The analysis of the intersection between the ISO trajectory and the ecliptic shows a peak in the range of 0.3-2 AU, with its maximum at $\sim$1 AU. This result shows that the Earth is placed at the optimum location to send spacecraft for ISO interception in its vicinity. This result also opens the door to the hypothesis that prebiotic molecules could have been brought to Earth by an ISO, since $\sim$1 AU is the location with a higher likelihood for ISO-ecliptic intersection. A high-level probability analysis has been performed and it has been determined that on average one ISO would hit the Earth every ten million years, leading to the hypothesis that the Earth could be the rare case where life was transmitted via an ISO. Although this hypothesis seems unlikely, the fact that the orbit of the Earth is in the range with the highest likelihood of ISO intersection should not be ignored. This shows that Earth is very well positioned to launch missions for ISO interception and highlights the potential of build-and-wait missions on Earth. If we take into account the effect of the difference in surface area, it is observed how orbits closer to the Sun have a higher probability.

Future work can improve the current understanding of these objects. Based on the analysis performed in this report, the author recommends to the space sector the investment of more financial resources in improving the current telescope surveying capabilities for ISO detection. Further work in confirming the results obtained in this project is also encouraged.

\graphicspath{ {images/} }
\justifying
\vspace{0mm}









\section*{\textbf{Appendix A: Build-and-wait Missions Trade-off}}

Build-and-wait missions, often called “launch on detection”, consist of building a spacecraft before the detection of the mission target, in this case, an ISO. This spacecraft can be kept on Earth where it would be stored in a clean room until an ISO is detected. After its detection, the spacecraft would be launched to intercept the target. Another possibility for the spacecraft is to be sent to space in a parking orbit where it would wait until a suitable target is detected. In this case, once the ISO is detected, the spacecraft would maneuver to start an intercept trajectory with the target. 

An example of a build-and-wait mission will be ESA’s Comet Interceptor spacecraft, expected to be launched in 2029. This spacecraft will travel to the Earth-Sun second Lagrange point, where it will wait to intercept a long period comet from the outer Solar System. Although more unlikely, this mission could also target an ISO if it was detected during the lifetime of the spacecraft \cite{Gater}.

In order to analyze the characteristics of both build-and-wait mission approaches and determine which one is better, a trade-off has been performed. 
For this reason, only parameters that are considered to be different in both cases have been assessed. Spacecraft manufacture, design and development have been assumed to be equal for both types of missions. The factors considered for the trade-off are cost, response time, risk, orbital maneuvers and spacecraft lifetime.
Note that the results of the trade-off analysis could change in the prospective future where the logistical aspects of mission launch could be improved. Table 1 shows the trade-off performed between build-and-wait missions on Earth and build-and-wait missions in space.


\begin{table*}[t]
\begin{center}
\caption{Trade-off analysis between build-and-wait missions on Earth and in space.}
\label{table:1}
\begin{tabular}{ |c|c|c|c| } 
\hline
 & \textbf{Buil-and-wait missions on Earth} & \textbf{Buil-and-wait missions on space} \\
\hline
   &
 Ready-to-use launch vehicle & Launch vehicle \\ 
  \textbf{Cost} &
 Maintenance and operations personnel
 & Operations personnel \\ 
  &
 Infrastructure for storage
 & Added in-orbit extra maneuvring technology \\ 
 \hline
 \textbf{Response time}  & Undetermined  & Immediate  \\
 \hline
 \textbf{Risk}  & Medium  & High \\
 \hline
 \textbf{Orbital}  & ISO interception  & Parking orbit maneuvres \\
  \textbf{maneuvres}& maneuvres & ISO interception maneuvres \\
  \hline
\textbf{Spacecraft}  & Decreased due to fuel limit and exposure  & Decreased due to fuel limit, prolonged exposure to \\
\textbf{lifetime}  & to space weather during ISO interception   & space weather during waiting time and ISO interception \\

\hline

\end{tabular}

\end{center}
\end{table*}

\begin{itemize}
    \item \textbf{Cost}
\end{itemize}
For both types of missions, a launch vehicle is required. The difference is, however, that build-and-wait missions on Earth require to have a ready-to-use launch vehicle prepared to ensure an adequate time response for the interception of the ISO. This could be accomplished by having a special arrangement with a launching company, which will induce a higher premium. Another option is the storage of a ready-to-use launch vehicle specifically for the ISO spacecraft. However, storing a ready-to-use launch vehicle would mean a great increase in the cost of the mission and, in addition, would not provide easy access to the launchpad, which would need to be arranged anyway with a launching company. It should be taken into account that many years may pass before the detection of a potential target. According to \cite{Seligman}, “wait times of the order of 10 years between favorable mission opportunities” are expected. If the mission cost wants to be minimized, the ISO needs to be intercepted in the vicinity of the Earth. Consequently, for build-and-wait missions on Earth, ready-to-use launch vehicles seem the most adequate, although expensive, option for this purpose. 
In addition, in build-and-wait missions on Earth, an infrastructure is required to store the spacecraft in order to have it in a launch-ready state. In this mission approach, personnel would be required to take care of the maintenance, system tests and updates that the stored spacecraft would require to ensure its proper functioning when launched. 

On the other hand, for build-and-wait missions in space, the date of the spacecraft launch is known in advance and, therefore, a preliminary cost analysis can be performed to decrease the launch vehicle cost. In addition, once the spacecraft is in space, personnel operating the spacecraft and taking care of its maintenance would be required. Furthermore, since in this mission approach the spacecraft would remain in a parking orbit, it would require additional technology and systems compared to build-and-wait missions on Earth. An example would be the incorporation of, according to \cite{Moore}, a solid rocket motor that would be required to enter an intercept trajectory with the target. 

In conclusion, build-and-wait missions on Earth have been considered to present a higher cost. This is because they would require to have a ready-to-use launch vehicle stored, and a special arrangement with a launching company to ensure fast response time.

\begin{itemize}
    \item \textbf{Response Time}
\end{itemize}

Due to the late detection and high heliocentric velocities that ISOs present, the window between detection and optimal trajectory interception in the vicinity of the Earth is short. Therefore, response time is essential.  
In a build-and-wait mission in space, the spacecraft would be in a stable orbit and ready to intercept the ISO when optimal. However, in the case of having a spacecraft waiting on Earth, it would be required to have the spacecraft prepared for launch, rapid launch capabilities and a ready-to-use launch vehicle. 
This supposes a great logistical challenge due to the organization of launches, as well as all the administrative and policy procedures. In this case, the response time may vary according to the financial capabilities of the company developing the mission and the special arrangements between the mission and launching company. 

An optimistic value for the response time based on the rapid response space program of the Operationally Responsive Space office would be of the order of days-months \cite{space}. However, it should be taken into account that this initiative is focused on developing small satellites and small launch vehicles for Earth orbit space capabilities. An estimated conservative value for the response time would be of the order of decades. The author of this report acknowledges the lack of a complete review of launch capabilities and frequencies. Thus, the author recommends the space sector to perform such review in order to have better estimates of response time, which is essential for these types of missions. 
The process of finding available space in an adequate launch vehicle with an appropriate launch window would decrease the response time. 

\begin{itemize}
    \item \textbf{Risk}
\end{itemize}

Build-and-wait missions to ISOs have high risks due to the fact that their detection is not ensured and, until now, only two of these objects have been identified. Designing a spacecraft for in-situ exploration beforehand is a challenge since the characteristics of the interstellar target are unknown. 

Spacecraft malfunctioning that can arise over the waiting years might suppose a high risk for the mission. In the case of build-and-wait missions on Earth, technical personnel would be able to control the state of the spacecraft technology and systems while it is stored. However, if the spacecraft is placed in space, it will be exposed to an extreme environment under the conditions of space weather. If waiting times are long, the radiation might damage the spacecraft components and the possibility of malfunctioning would increase. In these cases, the replacement of the spacecraft components in orbit would be difficult and would increase the mission cost. On the other hand, there are defined radiation risk mitigation alternatives for build-and-wait missions in space that could decrease the radiation effects. In addition, another mitigation strategy could be building a high degree of spacecraft autonomy to decrease the number of operations needed from Earth during the spacecraft waiting time in space. 

A common risk for both types of missions is the unknown precise location of the interstellar target, which is categorized as medium risk in Table 1 because it has a medium likelihood of occurrence. Since the ISO could present outgassing, the potential low precision on its location would be a risk to the mission. 
Furthermore, it should also be taken into account that build-and-wait missions in space might require additional motors and technology since the spacecraft needs to ignite in orbit upon target detection.

After having discussed these parameters, it can be observed that even though both mission approaches have different risks, build-and-wait missions in space present a higher risk due to the longer exposure to the space environment and its higher amount of maneuver operations. 

\begin{itemize}
    \item \textbf{Orbital Maneuvers}
\end{itemize}

Regarding orbital maneuvers, placing a spacecraft in a certain location in space and tracing a collision trajectory to the ISO would require more maneuvers than a direct interception trajectory. In addition, the use of fewer orbital maneuvers would result in less fuel required. For this reason, a spacecraft stationed in space would need more fuel than a spacecraft stored on Earth. This is due to the maneuvers needed to be performed to reach the parking orbit and its ISO interception trajectory maneuvers. 

\begin{itemize}
    \item \textbf{Spacecraft lifetime}
\end{itemize}

As it has been stated before, for build-and-wait missions in space, more maneuvers would be needed. Less fuel available for the spacecraft would reduce the waiting time to find a suitable interstellar target. Therefore, the lifetime of the spacecraft would be reduced. In addition, the constant exposure of the spacecraft to space weather would increase the radiation exposure of the system components, decrease its lifetime and, consequently, increase its risk of failure. The radiation exposure and temperature gradients would depend on the spacecraft distance and position from the Sun, its materials for radiation protection and the exposure time. As an example, the spacecraft of the Comet Interceptor mission has an expected lifetime of 5 years \cite{ESA}. 

On the other hand, build-and-wait missions on Earth would be launched for direct ISO interception. In this case, since the spacecraft would be stored on Earth, it would not be exposed to space weather during its waiting period.

With the aforementioned analysis, it has been observed that build-and-wait missions on Earth would present a longer spacecraft lifetime compared to build-and-wait missions in space.
Since both types of missions present high risk in mission success and a certain fixed lifetime, the author advises future build-and-wait missions to be multipurpose. Hence, if a potential ISO target has not been identified before the spacecraft lifetime ends, the spacecraft can be launched to study another celestial body.

\subsection{Results}
After discussing these different parameters, it has been concluded that build-and-wait missions on Earth are better than build-and-wait missions in space for the following parameters: orbital maneuvers, spacecraft lifetime and risk. However, build-and-wait missions in space have been determined to be better in the cost and response time parameters. 

The launcher response time and the special arrangements needed with the launching company to ensure fast response time have been considered a key factor to determine the outcome of the trade-off. Different estimations about the response time have been given, including an optimistic response time value of the order of days-months and an estimated conservative value of the order of decades. 

Both missions present different risks, which have also been considered one of the main factors to determine the outcome of the trade-off. However, build-and-wait missions in space present a higher risk due to the longer exposure to the space environment and their higher amount of maneuver operations. The lifetime of the spacecraft is also dependent on these factors. 

The author believes that build-and-wait missions on Earth are the best approach because of the risk and spacecraft lifetime analysis, its flexibility regarding spacecraft access and control while in storage. Nevertheless, with an increase in detection capability and the use of known mitigation strategies to decrease the space weather effects, build-and-wait missions in space stand as an appealing mission approach.
To make a final decision between these two types of missions, the future growth of immediate access to space launch capabilities should be considered. In 2020, 1085 spacecraft have been launched and 85 orbital launches have been performed, 12 of which were commercial and 73 non-commercial \cite{BryceTech}. The number of spacecraft launches has increased over the years and it is expected to keep increasing. Therefore, in the prospective future where access to space becomes even more accessible, rapid launch capabilities and improved detection capabilities would ensure the interception of an ISO with build-and-wait missions on Earth. In conclusion, the author believes that build-and-wait missions on Earth are the best mission approach for ISOs.

\section*{\textbf{Appendix B: Nomenclature}}
\raggedright
 
    $(a, b)$ =	Fehlberg coefficients \\
    $d$ = ISO distance from the origin of coordinates \\
    $e$ = Eccentricity \\
    $\tilde{f}$ = Derivative of Fehlberg’s Runge-Kutta method \\
    $f(x)$ = Function \\
    $G_c$ = Gravitational constant \\
     $M_1$ = Sun mass\\
    $M_2$ = ISO mass\\
    $O_s$ = Origin of coordinates\\
    $r$ = Radius\\
    $\vec{r}$ = Position vector\\
    $R_{Earth}$ = Earth radius \\
    $r_{inner}$ = Radius of the inner circumference\\
    $r_{outer}$ = Radius of the outer circumference \\
    $\vec{R}_{unit}$  = Unitary vector \\
    $\vec{R}_{1,0}$  = Initial position vector of the Sun \\
    $\vec{R}_{2,0}$  = Initial position vector of the ISO \\
    $r_2$ = ISO radius\\
    $t$ = Time variable of the Fehlberg’s Runge-Kutta method \\
     $\vec{v}$ = Velocity vector \\
    $|v_{LN}|$ = Magnitude of the Log-Normal distribution velocity\\
$v_r$ =  Radial velocity \\
$v_0$ = Initial velocity \\
$V_2$ = ISO volume \\
$\vec{v}_{2,0}$ = ISO initial velocity\\
$v_\infty$ =  Velocity at infinity\\
$(X_0,Y_0,Z_0)$ =Initial position vector in cartesian coordinates \\
$(X_1,Y_1,Z_1)$	= Position vector of the Sun in cartesian coordinates\\
$(X_2,Y_2,Z_2)$	= Position vector of an ISO in cartesian coordinates \\
$(\ddot{X}_1,\ddot{Y}_1,\ddot{Z}_1)$ =	Acceleration of the Sun in cartesian coordinates \\
$(\ddot{X}_2,\ddot{Y}_2,\ddot{Z}_2)$ = Acceleration of an ISO in cartesian coordinates \\
 $\alpha$ = Right ascension \\
$\delta$ = Declination \\
$\mu$ = Gravitational parameter\\
$\mu_{LN}$ = Mean of logarithmic values of the Log-Normal distribution\\
$\rho_2$ = ISO density \\
$\sigma$ =  Cross-section\\
$\sigma_{Earth}$ =  Cross-section of the Earth\\
$\sigma_{LN}$ = Standard deviation of logarithmic values of the Log-Normal distribution \\

\end{document}